\documentclass[preprint,showpacs,preprintnumbers,amsmath,amssymb]{revtex4}

\usepackage{graphicx}% Include figure files
\usepackage{dcolumn}% Align table columns on decimal point
\usepackage{bm}% bold math
\input{epsf}

\newcommand{\st}[2]{\stackrel{\mbox{\tiny (#1)}}{#2}\hspace{-0.1cm}}

\def\apj #1 #2 #3 {#1, ApJ, {\bf #2}, #3}
\def\apjl #1 #2 #3 {#1, ApJ, {\bf #2}, L#3}
\def\apjs #1 #2 #3 {#1, ApJS, {\bf #2}, #3}
\def\aap  #1 #2 #3 {#1, A\&A, {\bf #2}, #3}
\def\mnras #1 #2 #3 {#1, MNRAS, {\bf #2}, #3}
\def\pra #1 #2 #3 {#1, Phys.~Rev.~A., {\bf #2}, #3}
\def\prb #1 #2 #3 {#1, Phys.~Rev.~B., {\bf #2}, #3}
\def\prc #1 #2 #3 {#1, Phys.~Rev.~C., {\bf #2}, #3}
\def\prd #1 #2 #3 {#1, Phys.~Rev.~D., {\bf #2}, #3}
\def\pre #1 #2 #3 {#1, Phys.~Rev.~E., {\bf #2}, #3}
\def\prl #1 #2 #3 {#1, Phys.~Rev.~Lett., {\bf #2}, #3}
\def\plb #1 #2 #3 {#1, Phys.~Lett.~B., {\bf #2}, #3}
\def\science #1 #2 #3 {#1, Science., {\bf #2}, #3}
\def\nature #1 #2 #3 {#1, Nature., {\bf #2}, #3}
\def\nphysa #1 #2 #3 {#1, Nucl.~Phys.~A., {\bf #2}, #3}
\def\nphysb #1 #2 #3 {#1, Nucl.~Phys.~B., {\bf #2}, #3}
\def\nphysbs #1 #2 #3 {#1, Nucl.~Phys.~B.~Suppl., {\bf #2}, #3}

\def\h#1{\hbox{${}^{#1}$H}}

\def\h502{\hbox{$ h^{2}_{50}$}}

\def\la{\mathrel{\mathpalette\fun <}}
\def\ga{\mathrel{\mathpalette\fun >}}
\def\fun#1#2{\lower3.6pt\vbox{\baselineskip0pt\lineskip.9pt
  \ialign{$\mathsurround=0pt#1\hfil##\hfil$\crcr#2\crcr\sim\crcr}}}

\newcommand{\adota}{\frac{\dot a}{a}}
\newcommand{\vct}[1]{\mbox{\boldmath${#1}$}}
\begin{document}
\preprint{NTU-ASTRO-01}

\title{%
Primordial Neutrinos, Cosmological Perturbations \\
in Interacting Dark-Energy Model: CMB and LSS
}%

\author{Kiyotomo Ichiki$^{1,2,3}$
\footnote{Email address: ichiki@resceu.s.u-tokyo.ac.jp}, 
Yong-Yeon Keum$^{3,4}$
\footnote{Email address: yykeum@phys.ntu.edu.tw} }

\affiliation{%
$^1$National Astronomical Observatory, Mitaka, Tokyo 181-8588, Japan
}%
\affiliation{%
$^2$Kavli Institute for Cosmological Physics, University of Chicago, IL 60637, USA
}%

\affiliation{%
$^3$Department of Physics, National Taiwan University, Taipei, Taiwan 10672, ROC}%

\affiliation{%
$^4$Asia Pacific Center for Theoretical Physics, Pohang, Korea}

\vskip1.0cm
%\date{\today}

\vskip1.0cm
\begin{abstract}
We present cosmological perturbation theory in neutrinos probe
interacting dark-energy models, and calculate cosmic microwave
background anisotropies and matter power spectrum. In these models,
the evolution of the mass of neutrinos is determined by the
quintessence scalar field, which is responsible for the cosmic
acceleration today. We consider several types of scalar field
potentials and put constraints on the coupling parameter between
neutrinos and dark energy. Assuming the flatness of the universe,
the constraint we can derive from the current observation is $\sum
m_{\nu} < 0.87 eV$ at the 95 $\%$ confidence level for the sum over
three species of neutrinos. We also discuss on the stability issue of 
the our model and on the impact of the scattering term in Boltzmann
equation from the mass-varying neutrinos.
\end{abstract}

\keywords{ Time Varying Neutrino Masses; Neutrino Mass Bound; Cosmic
Microwave Background; Large Scale Structures; Quintessence Scalar
Field}

\pacs{ 98.80.-k,98.80.Jk,98.80.Cq}

\thispagestyle{empty}
\maketitle
%%%%%%%%%%%%%%%%%%%%%%%%%%%%%%%%%%%%%%%%%%%%%%%%%%%%%%%%%%
\newpage

\section{Introduction}
After SNIa\cite{sn1a} and WMAP\cite{wmap} observations during last
decade, the discovery of the accelerated expansion of the universe
is a major challenge of particle physics and cosmology. There are
currently three candidates for the {\it Dark-Energy} which derives
this accelerated expansion:
\begin{itemize}
\item a non-zero cosmological constant\cite{lambda},
\item a dynamical cosmological constant (Quintessence scalar field)\cite{quintessence},
\item modifications of Einstein Theory of Gravity\cite{mgrav}
\end{itemize}
The scalar field model like quintessence
is a simple model with time dependent $w$, which is generally larger
than $-1$. Because the different $w$ leads to a different expansion
history of the universe, the geometrical measurements of  
cosmic expansion through observations of SNIa, CMB, and Baryon Acoustic
Oscillations (BAO) can give us tight
constraints on $w$.
One of the interesting way to study the scalar field dark energy models is
to investigate the coupling between the dark energy and the other matter
fields. In fact, a number of models which realize the interaction
between dark energy and dark matter, or even visible matters, have been
proposed so far
\cite{Carroll:1998zi,Bean:2000zm,Farrar:2003uw,das-khoury:2006,Lee:2006za}. 
Observations of the effects of these interactions will offer an unique
opportunity to detect a cosmological scalar field
\cite{Carroll:1998zi,Liu:2006uh}. 

In this paper, after reviewing shortly the main idea of the three
possible candidates of dark energy and their cosmological phenomena in
section II, we discuss the interacting dark energy model, paying
particular attention to the interacting mechanism between
dark-energy with a hot dark-matter (neutrinos) in section III. In this
so-called Mass-Varying Neutinos (MaVaNs) model \cite{mavanu}, we
calculate explicitly Cosmic Microwave Background(CMB) radiation and
Large Scale Structure(LSS) within cosmological perturbation theory. 
The evolution of the mass of neutrinos is determined by the quintessence
scalar filed, which is responsible for the cosmic acceleration
today. Recently, perturbation equations for this class of models are
nicely presented by Brookfield et al. \cite{Brookfield-b}, (see also
\cite{Zhao:2006zf}) which are necessary to compute CMB and LSS
spectra. A main difference here from their works is that we correctly
take into account the scattering term in the geodesic equation of
neutrinos, which was omitted there (see, however, \cite{Brookfield:2005bz}).
We will show that this leads significant
differences in the resultant spectra and hence the different
observational constraints. In section IV, we discuss three different
types of quintessence potential, namely, an inverse power law potential,
a supergravity potential, and an exponential type potential. By
computing CMB and LSS spectra with these quintessential potentials
and comparing them to the latest observations, the constraints on the
present mass of neutrinos and coupling parameters are derived. 
In conclusion we discuss two important points of this work 
on the impact of the scattering term of the Boltzmann equation 
and on the stability issue in the interacting dark-energy model. 
In appendix A, the explicit calculation for
the consistency check of our calculations in section III is shown.
Since we were asked to show explicit derivation of geodesic
equation after our first draft was released, we show them in 
appendix B.

\section{Three possible solutions for Accelerating Universe:}
Recent observations with Supernova Ia type (SNIa) and CMB radiation
have provided strong evidence that we live now in an accelerating
and almost flat universe. In general, one believes that the
dominance of a dark-energy component with negative pressure in the
present era is responsible for the universe's accelerated expansion.
However there are three possible solutions to explain the
accelerating universe. The Einstein Equation in General Relativity
is given by the following form:
\begin{equation}
G_{\mu\nu} \,\,= \,\, R_{\mu\nu} -{1 \over 2} R \, g_{\mu\nu} \hspace{2mm}
= \hspace{2mm} 8 \pi G \, T_{\mu\nu} \hspace{2mm} + \hspace{2mm} \Lambda \, g_{\mu\nu},
\label{eq:einstein}
\end{equation}
Here, $G_{\mu\nu}$ term contains the information of geometrical
structure, the energy-momentum tensor $T_{\mu\nu}$ keeps the
information of matter distributions, and the last term is so called
the cosmological constant which contain the information of non-zero
vacuum energy. After solve the Einstein equation, one can drive a
simple relation:
\begin{equation}
{\ddot{R} \over R} = -{4 \pi G \over 3} (\rho +3p) +{\Lambda \over
3}.
\end{equation}
In order to get the accelerating expansion, either cosmological
constant $\Lambda$ ($\omega_{\Lambda} = P/\rho=-1$) becomes positive
or a new concept of dark-energy with the negative pressure
($\omega_{\phi} <-1/3$) needs to be introduced. Another solution can
be given by the modification of geometrical structure which can
provide a repulsive source of gravitational force.  In this case,
the attractive gravitational force term is dominant in early stage
of universe, however at later time near the present era, repulsive
term become important and drives universe to be expanded with an
acceleration.
 Also we can consider extra-energy density
contributions from bulk space in Brane-World scenario models, which
can modify the Friedmann equation as $H^2 \propto \rho + \rho^{'}$.
In summary, we have three different solutions for the accelerating
expansion of our universe as mentioned in the introduction.
 Probing for the origin of accelerating universe is the most
important and challenged problem in high energy physics and
cosmology now. The detail explanation and many references are in a
useful review on dark energy\cite{review-DE}.

In this paper, we concentrate on the second solution using the
quintessence field. In present epoch, the potential term becomes
important than kinetic term, which can easily explain the negative
pressure with $\omega_{\phi}^0 \simeq -1$. However there are many
different versions of quintessence field: K-essence\cite{COY,k-essence},
phantom\cite{phantom}, quintom\cite{quintom}, ....etc., and to
justify the origin of dark-energy from experimental observations is
really a difficult job. Present updated value of the equation of
states(EoS) are $\omega=-1.02 \pm 0.12$ without any supernova
data\cite{seljak:0604335}.

\section{Interacting Dark-Energy with Neutrinos:}
As explained in previous section, it is really difficult to probe
the origin of dark-energy when the dark-energy doesn't interact with
other matters at all. Here we investigate the cosmological
implication of an idea of the dark-energy interacting with neutrinos
\cite{{mavanu},{Fardon:2003eh}}. For simplicity, we consider the
case that dark-energy and neutrinos are coupled such that the mass
of the neutrinos is a function of the scalar field which drives the
late time accelerated expansion of the universe. In previous works
by Fardon et al.\cite{Fardon:2003eh} and R. Peccei\cite{mavanu},
kinetic energy term was ignored and potential term was treated as a
dynamical cosmology constant, which can be applicable for the
dynamics near present epoch. However the kinetic contributions
become important to descreibe cosmological perturbations in early
stage of universe, which is fully considered in our analysis.

\subsection{Cosmological perturbations: background Equations}
Equations for quintessence scalar field are given by
\begin{eqnarray}
\ddot{\phi}&+&2{\cal H}\dot\phi+a^2\frac{d V_{\rm eff}(\phi)}{d\phi}=0~,
 \label{eq:Qddot}\\
V_{\rm eff}(\phi)&=&V(\phi)+V_{\rm I}(\phi)~,\\
V_{\rm I}(\phi)&=&a^{-4}\int\frac{d^3q}{(2\pi)^3}\sqrt{q^2+a^2
 m_\nu^2(\phi)}f(q)~,\\
m_\nu(\phi) &=& \bar m_i e^{\beta\frac{\phi}{M_{\rm pl}}}~{\mbox{(as an example)}},
\end{eqnarray}
where $V(\phi)$ is the potential of quintessence scalar field, $V_{\rm
I}(\phi)$ is additional potential due to the coupling to neutrino
particles \cite{Fardon:2003eh,Bi:2003yr},
and $m_\nu(\phi)$ is the mass of neutrino coupled to the scalar field.
${\cal H}$ is $\frac{\dot a}{a}$, where the dot represents the
derivative with respect to the conformal time $\tau$.

Energy densities of mass varying neutrino (MVN) and quintessence scalar
field are described as
\begin{eqnarray}
\rho_\nu &=& a^{-4}\int \frac{d^3 q}{(2\pi)^3} \sqrt{q^2+a^2m_\nu^2} f_0(q)~, \label{eq:rho_nu}\\
3P_\nu &=& a^{-4}\int \frac{d^3 q}{(2\pi)^3} \frac{q^2}{\sqrt{q^2+a^2m_\nu^2}}
 f_0(q)~, \label{eq:P_nu}\\
\rho_\phi &=& \frac{1}{2a^2}\dot\phi^2+V(\phi)~,\\
P_\phi &=& \frac{1}{2a^2}\dot\phi^2-V(\phi)~.
\end{eqnarray}
From equations (\ref{eq:rho_nu}) and (\ref{eq:P_nu}), the equation of
motion for the background energy density of neutrinos is given by
\begin{equation}
\dot\rho_{\nu}+3{\cal H}(\rho_\nu+P_\nu)=\frac{\partial \ln
 m_\nu}{\partial \phi}\dot\phi(\rho_\nu -3P_\nu)~.
\end{equation}

\subsection{Perturbation equations}
\subsubsection{perturbations in the metric}
We work in the synchronous gauge and line element is
\begin{equation}
 ds^2 = a^2(\tau)\left[-d\tau^2 + (\delta_{ij}+h_{ij})dx^i dx^j\right]~,
\end{equation}
In this metric the Chirstoffel symbols which have non-zero values are
\begin{eqnarray}
\Gamma^0_{00}&=&\adota~, \\
\Gamma^0_{ij}&=&\adota \delta_{ij}+\frac{\dot
 a}{a}h_{ij}+\frac{1}{2}\dot h_{ij}~, \\
\Gamma^i_{0j}&=&\adota\delta^i_j+\frac{1}{2}\dot h_{ij}~,\\
\Gamma^i_{jk}&=&\frac{1}{2}\delta^{ia}(h_{ka,j}+h_{aj,k}-h_{jk,a})~,
\end{eqnarray}
where dot denotes conformal time derivative.
For CMB anistropies we mainly consider the scalar type perturbations. We
introduce two scalar fields, $h(\vct k, \tau)$ and $\eta(\vct k, \tau)$, in
k-space and write the scalar mode of $h_{ij}$ as a Fourier integral
\cite{Ma:1995ey}
\begin{equation}
h_{ij}(\vct x,\tau)=\int d^3k e^{i\vct{k}\cdot\vct{x}}\left[\vct{\hat
 k}_i\vct{\hat k}_j h(\vct k,\tau)+(\vct{\hat k}_i \vct{\hat k}_j
 -\frac{1}{3}\delta_{ij})6\eta(\vct{k},\tau) \right]~,
\end{equation}
where $\vct{k} =k\vct{\hat k}$ with $\hat{k}^i \hat{k}_i =1$.
\subsubsection{perturbations in quintessence}
The equation of quintessence scalar field is given by
\begin{equation}
 \Box \phi - V_{\rm eff}(\phi)=0~.
\end{equation}
Let us write the scalar field as a sum of background value and
perturbations around it, $\phi(\vct x,\tau)=\phi(\tau)+\delta \phi(\vct
x,\tau)$.
The perturbation equation is then described as
\begin{equation}
\frac{1}{a^2}\ddot{\delta \phi}+\frac{2}{a^2}{\cal H}\dot\delta\phi -
 \frac{1}{a^2}\nabla^2(\delta\phi)+\frac{1}{2a^2}\dot{h}\dot\phi+\frac{d^2
 V}{d\phi^2}\delta\phi +\delta\left(\frac{dV_{\rm I}}{d\phi}\right)=0~ \label{eq:dQddot},
\end{equation}
where
\begin{eqnarray}
\frac{d V_{\rm I}}{d\phi} &=& a^{-4}\int
 \frac{d^3q}{(2\pi)^3}\frac{\partial \epsilon(q,\phi)}{\partial \phi}
 f(q)~,\label{eq:dvidphi}\\
\epsilon(q,\phi) &=& \sqrt{q^2+a^2 m_\nu^2(\phi)} ~,\\
\frac{\partial \epsilon(q,\phi)}{\partial \phi} &=& \frac{a^2
 m^2_\nu(\phi)}{\epsilon(q,\phi)} \frac{\partial \ln m_\nu}{\partial \phi}~.
\end{eqnarray}
To describe $\delta\left(\frac{dV_{\rm I}}{d\phi}\right)$,
we shall write the distribution function of neutrinos with background
distribution and perturbation around it as
\begin{equation}
f(x^i, \tau, q, n_j )= f_0(\tau,q)(1+\Psi(x^i, \tau, q, n_j))~.
\end{equation}
Then we can write
\begin{equation}
\delta\left(\frac{dV_{\rm I}}{d\phi}\right)
 =a^{-4}\int\frac{d^3q}{(2\pi)^3}
 \frac{\partial^2 \epsilon}{\partial \phi^2}\delta\phi f_0
 + a^{-4}\int\frac{d^3q}{(2\pi)^3}\frac{\partial \epsilon}{\partial
 \phi}f_0\Psi~, \label{eq:delta(dvidphi)}
\end{equation}
where
\begin{eqnarray}
\frac{\partial^2 \epsilon}{\partial \phi^2}&=&\frac{a^2}{\epsilon}\left(\frac{\partial m_\nu}{\partial \phi}\right)^2 +\frac{a^2 m_\nu}{\epsilon}\left(\frac{\partial^2 m_\nu}{\partial \phi^2}\right) \nonumber \\
&-&\frac{a^2 m_\nu}{\epsilon^2}\left(\frac{\partial \epsilon}{\partial \phi}\right)\left(\frac{\partial m_\nu}{\partial \phi}\right)~.
\end{eqnarray}
For numerical purpose it is useful to rewrite the equations
(\ref{eq:dvidphi}) and (\ref{eq:delta(dvidphi)}) as
\begin{eqnarray}
\frac{dV_{\rm I}}{d\phi} &=& \frac{\partial \ln m_\nu}{\partial
 \phi}(\rho_\nu -3P_\nu)~, \\
\delta\left(\frac{dV_{\rm I}}{d\phi}\right) &=&
\frac{\partial^2 \ln m_\nu}{\partial
 \phi^2} \delta\phi (\rho_\nu-3P_\nu) \nonumber \\
&&+ \frac{\partial \ln m_\nu}{\partial
 \phi}(\delta\rho_\nu-3\delta P_\nu)
\end{eqnarray}
%where $\Theta$ is defined as
%\begin{equation}
%\Theta = a^{-4}\int \frac{d^3 q}{(2\pi)^3}\frac{a^2 m^2
% q^2}{\epsilon^3}f_0~,
%\end{equation}
Note that perturbation fluid variables in mass varying neutrinos are given by
\begin{eqnarray}
\delta\rho_\nu&=&a^{-4}\int\frac{d^3 q}{(2\pi)^3} \epsilon
 f_0(q)\Psi+a^{-4}\int\frac{d^3 q}{(2\pi)^3}\frac{\partial
 \epsilon}{\partial \phi}\delta\phi f_0 ~, \label{eq:delta_rho_nu}\\
3\delta P_\nu&=&a^{-4}\int\frac{d^3 q}{(2\pi)^3} \frac{q^2}{\epsilon}
 f_0(q)\Psi-a^{-4}\int\frac{d^3
 q}{(2\pi)^3}\frac{q^2}{\epsilon^2}\frac{\partial \epsilon}{\partial
 \phi}\delta\phi f_0 ~.
\end{eqnarray}
The energy momentum tensor of quintessence is given by
\begin{equation}
 T^\mu_\nu = g^{\mu\alpha}\phi_{,\alpha}\phi_{,\nu}-\frac{1}{2}\left(\phi^{,\alpha}\phi_{,\alpha}+2V(\phi)\right)\delta^\mu_\nu~,
\end{equation}
and its perturbation is
\begin{eqnarray}
 \delta T^\mu_\nu&=&g_{(0)}^{\mu \alpha}\delta\phi_{,\alpha}\phi_{,\nu}
  +g_{(0)}^{\mu \alpha}\phi_{,\alpha}\delta \phi_{,\nu}
  +\delta g^{\mu \alpha}\phi_{,\alpha}\phi_{,\nu}\nonumber \\
  &-&\frac{1}{2}\left(\delta\phi^{,\alpha}\phi_{,\alpha}
          +\phi^{,\alpha}\delta\phi_{,\alpha}
          +2\frac{dV}{d\phi}\delta\phi\right)\delta^\mu_\nu~.
\end{eqnarray}
This gives perturbations of quintessence in fulid variables as
\begin{eqnarray}
 \delta\rho_\phi&=&-\delta T^0_0 =
 \frac{1}{a^2}\dot\phi\dot{\delta\phi}+\frac{dV}{d\phi}\delta\phi~, \\
 \delta P_\phi&=&-\delta T^0_0/3 =
 \frac{1}{a^2}\dot\phi\dot{\delta\phi}-\frac{dV}{d\phi}\delta\phi~, \\
 (\rho_\phi +P_\phi)\theta_\phi&=&ik^i\delta
 T^0_i=\frac{k^2}{a^2}\dot\phi\delta\phi~,\\
 \Sigma^i_j&=&T^i_j-\delta^i_j T^k_k/3=0~.
\end{eqnarray}

\subsection{Boltzmann Equation for Mass Varying Neutrino}
We have to consider Boltzmann equation to solve the evolution of VMN.
A distribution function is written in terms of time
($\tau$), positions ($x^i$) and their conjugate momentum ($P_i$).
The conjugete momentum is defined as spatial parts of the 4-momentum
with lower indices, i.e., $P_i = mU_i$, where
$U_i=dx_i/(-ds^2)^{1/2}$. We also introduce locally orthonormal
coordinate $X^\mu = (t,r^i)$, and we write the energy and the momentum
in this coordinate as $(E,p^i)$, where $E=\sqrt{p^2+m_\nu^2}$.
The relations of these variables in synchronous gauge are given by
\cite{Ma:1995ey},
\begin{eqnarray}
 P_0 &=& -aE~, \\
 P_i &=& a(\delta_{ij}+\frac{1}{2}h_{ij})p^{j}~.
\end{eqnarray}
Next we define comoving energy and momentum $(\epsilon, q_i)$ as
\begin{eqnarray}
 \epsilon &=& aE = \sqrt{q^2+a^2 m_\nu^2}~,\\
 q_i &=& ap_i~.
\end{eqnarray}
Hereafter, we shall use ($x^i,q,n_j,\tau$) as phase space variables,
replacing $f(x^i,P_j,\tau)$ by $f(x^i,q,n_j,\tau)$. Here
we have splitted the comoving momentum $q_j$ into its magnitude and
direction: $q_j = qn_j$, where $n^i n_i=1$.
The Boltzmann equation is
\begin{equation}
 \frac{Df}{D\tau}=\frac{\partial f}{\partial
 \tau}+ \frac{dx^i}{d\tau}\frac{\partial f}{\partial
 x^i}
+\frac{dq}{d\tau}\frac{\partial f}{\partial
 q}+\frac{dn_i}{d\tau}\frac{\partial f}{\partial
 n_i}=\left(\frac{\partial f}{\partial \tau}\right)_C~.
\end{equation}
in terms of these variables.
From the time component of geodesic equation \cite{Anderson:1997un},
\begin{equation}
 \frac{1}{2}\frac{d}{d\tau}\left(P^0\right)^2=-\Gamma^0_{\alpha\beta}P^\alpha
  P^\beta - m g^{0 \nu}m_{,\nu}~,
\label{eq:eq-a}
\end{equation}
and the relation $P^0=a^{-2}\epsilon=a^{-2}\sqrt{q^2+a^2 m_\nu^2}$, we
have
\begin{equation}
 \frac{dq}{d\tau}=-\frac{1}{2}\dot{h_{ij}}qn^in^j-a^2
  \frac{m}{q}\frac{\partial m}{\partial x^i}\frac{dx^i}{d \tau} ~.
\label{eq:eq-b}
\end{equation}
Our analytic formulas in eqs.(\ref{eq:eq-a}-\ref{eq:eq-b}) are different
from those of \cite{Brookfield-b} and \cite{Zhao:2006zf}, since they
have omitted the contribution of the varying neutrino mass term. We
shall show later this term also give an important contribution in the
first order perturbation of the Boltzman equation. 

We will write down each term up to ${\cal O}(h)$:
\begin{eqnarray}
 \frac{\partial f}{\partial \tau} &=&\frac{\partial f_0}{\partial
 \tau}+f_0 \frac{\partial \Psi}{\partial \tau} +\frac{\partial
 f_0}{\partial \tau}\Psi \nonumber \\
 \frac{dx^i}{d\tau}\frac{\partial f}{\partial
 x^i}&=&\frac{q}{\epsilon}n^i \times f_0\frac{\partial \Psi}{\partial
 x^i} ~,\nonumber \\
 \frac{dq}{d\tau}\frac{\partial f}{\partial
 q}&=&\left(-a^2\frac{m_\nu}{q}\frac{\partial m_\nu}{\partial x^i}\frac{dx^i}{d\tau}-\frac{1}{2}\dot{h_{ij}}qn^in^j\right)
\times\frac{\partial f_0}{\partial q} \nonumber \\
 \frac{dn_i}{d\tau}\frac{\partial f}{\partial
 n_i}&=& {\cal O}(h^2)~.
\end{eqnarray}
We note that $\frac{\partial f}{\partial
 x^i}$ and $\frac{dq}{d\tau}$ are ${\cal O}(h)$.

\subsubsection{Background equations}
From the equations above, the zeroth-order Boltzmann equation is
\begin{equation}
 \frac{\partial f_0}{\partial\tau}=0~.
\label{eq:Boltz0}
\end{equation}
The Fermi-Dirac distribution
\begin{equation}
 f_0=f_0(\epsilon)=\frac{g_s}{h_{\rm P}^3}\frac{1}{e^{\epsilon/k_{\rm B}T_0}+1}~,
\end{equation}
can be a solution.
Here $g_s$ is the number of spin degrees of freedom, $h_{\rm P}$ and
$k_{\rm B}$ are the Planck and the Boltzmann constants. 

\subsubsection{perturbation equations}
The first-order Boltzmann equation is
\begin{eqnarray}
 \frac{\partial \Psi}{\partial
 \tau}&+&i\frac{q}{\epsilon}(\vct{\hat{n}}\cdot\vct{k})\Psi+\left(\dot\eta-(\vct{\hat
 k}\cdot\vct{\hat n})^2\frac{\dot h+6\dot\eta}{2}\right)\frac{\partial \ln
 f_0}{\partial \ln q}\nonumber \\
&-&i\frac{q}{\epsilon}(\vct{\hat{n}}\cdot\vct{k})k\delta\phi\frac{a^2
 m^2}{q^2}\frac{\partial \ln m}{\partial \phi}\frac{\partial \ln
 f_0}{\partial \ln q} = 0~.
\label{eq:boltzmann}
\end{eqnarray}
Following previous studies, we shall assume that the initial momentum
dependence is axially symmetric so that $\Psi$ depends on
$\vct{q}=q\vct{\hat n}$ only through $q$ and $\vct{\hat k}\cdot\vct{\hat
n}$.
With this assumption, we expand the perturbation of distribution
function, $\Psi$, in a Legendre series,
\begin{equation}
 \Psi(\vct{k},\vct{\hat n},q,\tau)=\sum (-i)^\ell(2\ell+1)\Psi_\ell(\vct{k},q,\tau)P_\ell(\vct{\hat{k}}\cdot\vct{\hat{n}})~.
\end{equation}
Then we obtain the hierarchy for MVN
\begin{eqnarray}
 \dot{\Psi_0}&=&-\frac{q}{\epsilon}k\Psi_1
                +\frac{\dot h}{6}\frac{\partial \ln{f_0}}{\partial\ln{q}}~, \label{eq:dot_Psi_0}\\
\dot{\Psi_1}&=&\frac{1}{3}\frac{q}{\epsilon}k\left(\Psi_0-2\Psi_2\right)
              + \kappa~, \label{eq:dot_Psi_1}\\
\dot{\Psi_2}&=&\frac{1}{5}\frac{q}{\epsilon}k(2\Psi_1-3\Psi_3)-\left(\frac{1}{15}\dot{h}+\frac{2}{5}\dot{\eta}\right)\frac{\partial \ln{f_0}}{\partial \ln{q}}~,\\
\dot{\Psi_\ell}&=&\frac{q}{\epsilon}k\left(\frac{\ell}{2\ell+1}\Psi_{\ell-1}-\frac{\ell+1}{2\ell+1}\Psi_{\ell+1}\right)~.
\end{eqnarray}
where
% \begin{eqnarray}
% \gamma &=& \left\{\dot{\delta \phi}\frac{\partial \ln m_\nu}{\partial \phi}
%      + \dot\phi \delta\phi\left[\left(\frac{\partial \ln
%                      m_\nu}{\partial
%                      \phi}\right)^2+\frac{\partial^2
%      \ln m_\nu}{\partial \phi^2}\right]
%      \right\}\frac{a^2 m_\nu^2}{\epsilon^2} \label{eq:gamma}\nonumber \\
%  &&+\dot\phi \delta\phi\left(\frac{\partial \ln m_\nu}{\partial
%           \phi}\right)^2 \frac{a^2 m_\nu^2
%  q^2}{\epsilon^4}~, \\
% \kappa &=& \frac{q}{\epsilon} k \delta\phi\left(\frac{\partial \ln
%                      m_\nu}{\partial
%                      \phi}\right)\frac{a^2
% m_\nu^2}{q^2}
% \label{eq:kappa}
% \end{eqnarray}
\begin{equation}
\kappa = -\frac{1}{3}\frac{q}{\epsilon}k\frac{a^2 m^2}{q^2}\delta\phi
 \frac{\partial \ln m_\nu}{\partial \phi}\frac{\partial \ln
 f_0}{\partial \ln q}\label{eq:kappa}~.
\end{equation}
Here we used the recursion relation
\begin{equation}
 (\ell+1)P_{\ell+1}(\mu)=(2\ell+1)\mu P_\ell(\mu)-\ell P_{\ell-1}(\mu)~.
\end{equation}
We have to solve these equations with a $q$-grid for every wavenumber $k$.

\section{Quintessence potentials}
To determine the evolution of scalar field which couples to neutrinos, we
should specify the potential of the scalar field.
A variety of quintessence effective potentials can be found in the
literature.
In the present paper we examine three type of quintessential potentials.
First we analyze what is a frequently invoked form for the effective
potential of the tracker field, i.e., an inverse power law such as
originally analyzed by Ratra and Peebles \cite{Ratra:1987rm},
\begin{equation}
V(\phi)=M^{4+\alpha}\phi^{-\alpha}~~~\mbox{(Model I)}~,
\end{equation}
where $M$ and $\alpha$ are parameters.

We will also consider a modified form of $V(\phi)$ as proposed by
\cite{Brax:1999gp} based on the condition that the quintessence fields
be part of supergravity models. The potential now becomes
\begin{equation}
V(\phi)=M^{4+\alpha}\phi^{-\alpha} e^{3\phi^2/2m_{\rm pl}^2}~~~\mbox{(Model II)}~,
\end{equation}
where the exponential correction becomes important near the present time
as $\phi \to m_{\rm pl}$.
The fact that this potential has a minimum for
$\phi=\sqrt{\alpha/3}m_{\rm pl}$ changes the dynamics.
It causes the present value of $w$ to evolve to a cosmological constant
much quicker than for the bare power-law potential \cite{Brax:2000yb}.
In these models the parameter $M$ is fixed by the condition that
$\Omega_Q\approx 0.7$ at present.

\begin{figure}
    \rotatebox{0}{\includegraphics[width=0.60\textwidth]{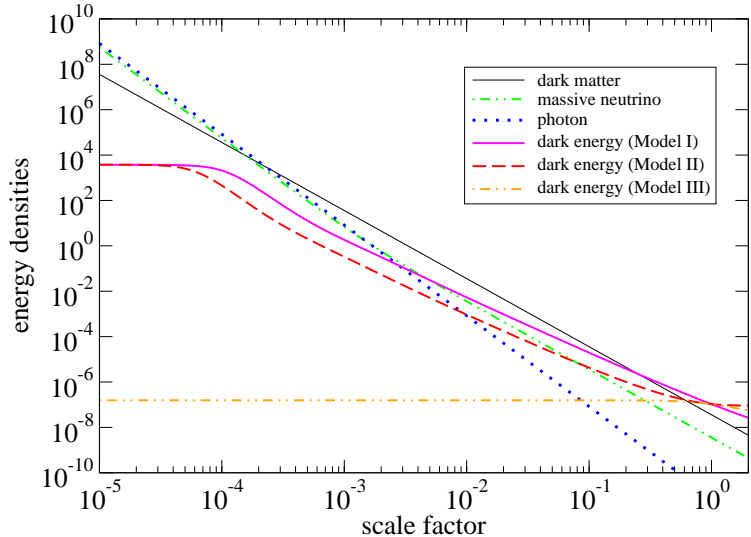}}
\caption{Examples of the evolution of energy density in quintessence and
 the background fields as indicated. Model parameters taken to plot this
 figure are $\alpha=10$, $10$, $1$ for model I, II, III,
 respectively. The other parameters for the dark energy are fixed so that
 the energy densities in three types of dark energy should be the same at
 present.
}
\label{fig:energy_densities}
\end{figure}

We will also analyze another class of tracking potential, namely, the
potential of exponential type \cite{Copeland:1997et}:
\begin{equation}
V(\phi) = M^4 e^{-\alpha\phi}~~~\mbox{(Model III)}~,
\end{equation}
This type of potential can lead to accelerating expansion provided that
$\alpha<\sqrt{2}$. In figure (\ref{fig:energy_densities}), we present
examples of evolution of
energy densities with these three types of potentials with vanishing
coupling strength to neutrinos.

\subsection{Time evolution of neutrino mass and energy density in scalar
 field}
\begin{figure}
    \rotatebox{0}{\includegraphics[width=0.60\textwidth]{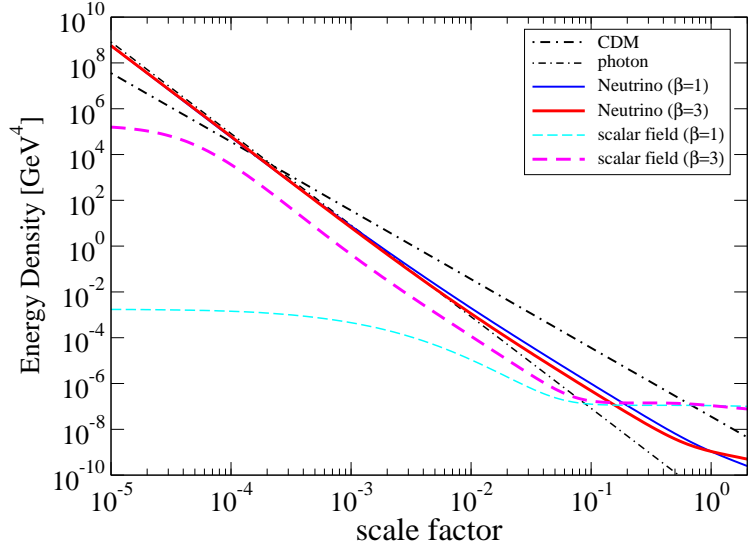}}
\caption{Examples of the evolution of energy density in quintessence and
 the background fields in coupled cases with inverse power law potential
 (Model I). Model parameters taken to plot this
 figure are $\alpha=1$, $\beta=1$, $3$ as indicated. The other
 parameters for the dark energy are fixed so that
 the energy densities in three types of dark energy should be the same at
 present.
}
\label{fig:energy_densities_coupled}
\end{figure}

\begin{figure}
    \rotatebox{0}{\includegraphics[width=0.60\textwidth]{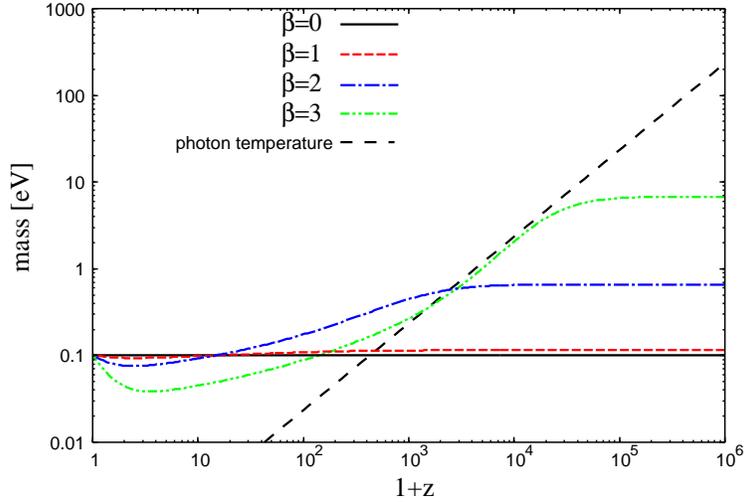}}
\caption{Examples of the time evolution of neutrino mass in power law
 potential models (Model I) with $\alpha=1$ and $\beta=0$ (black solid line),
 $\beta=1$ (red dashed line), $\beta=2$ (blue dash-dotted line),
 $\beta=3$ (dash-dot-dotted line). The larger coupling parameter leads
 to the larger mass in the early universe.
}
\label{fig:mass_evolution}
\end{figure}

For an illustration we also plot examples of evolution of energy
densities for interacting case with inverse power law potential (Model
I) in Fig. (\ref{fig:energy_densities_coupled}).
In interacting dark energy cases, the evolution of the scalar field is
determined both by its own potential and interacting term from neutrinos.
When neutrinos are highly relativistic, the interaction term can be expressed
as
\begin{equation}
\frac{\partial m_\nu}{\partial \phi}(\rho_\nu-3P_\nu)\approx
 \frac{10}{7\pi^2} (am_\nu)^2 \rho_{\nu_{\rm massless}}~,
\end{equation}
where $\rho_{\nu_{\rm massless}}$ denotes the energy density of neutrinos
with no mass.
The term roughly scales as $\propto a^{-2}$, and therefore, it dominates
deep in the radiation dominated era.
However, because the motion of the scalar field driven by this interaction term
is almost suppressed by the friction term, $-2{\cal H}\dot\phi$. The
scalar field satisfies the slow roll condition similar to the inflation
models, $-2{\cal H}\dot\phi\approx a^2\frac{\partial m_\nu}{\partial
\phi}(\rho_\nu-3P_\nu)$. Thus, the energy density in scalar field
and the mass of neutrinos is frozen there. These behaviors are clearly seen in
Figs. (\ref{fig:energy_densities_coupled}) and (\ref{fig:mass_evolution}).

\subsection{Constrains on the MaVaNs parameters}
As was shown in the previous sections, the coupling between cosmological
neutrinos and dark energy quintessence could modify the CMB and matter
power spectra significantly.  It is therefore possible and also
important to put
constraints on coupling parameters from current observations.
For this purpose, we use the WMAP3 \cite{Hinshaw:2006ia,Page:2006hz} and
2dF \cite{Cole:2005sx} data sets.

\begin{figure}
    \vspace*{10mm}\rotatebox{0}{\includegraphics[width=0.60\textwidth]{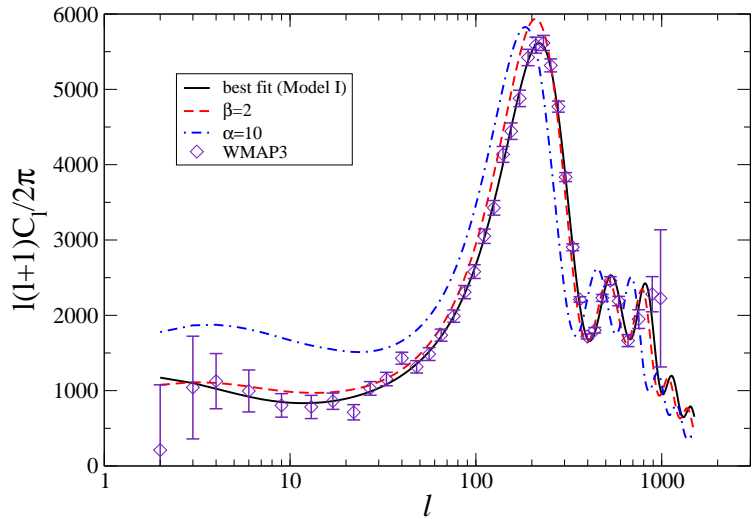}}
\caption{The CMB angular power spectra for Model I. The solid line is
 the best fit for the model ($(\alpha,\beta)=(2.97,0.170)$), the other lines are models with different
 parameter value of $\alpha$ and $\beta$ as indicated.
The points are WMAP three year data.
}
\label{fig:cl_spectra}
\end{figure}

\begin{figure}
    \rotatebox{0}{\includegraphics[width=0.60\textwidth]{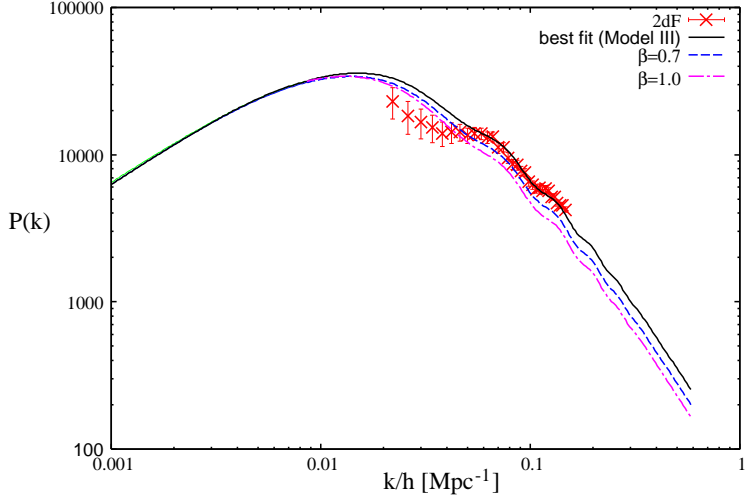}}
\caption{The CMB angular power spectra for Model III. The solid line is
 the best fit for the model ($(\alpha,\beta)=(0.78,0.28)$), the other
 lines are models with different
 parameter value of $\alpha$ and $\beta$ as indicated.
The points are 2dF data.
}
\label{fig:pk_spectra}
\end{figure}

The flux power spectrum of the Lyman-$\alpha$ forest can be used to
measure the matter power spectrum at small scales around $z\la 3$ \cite{McDonald:1999dt,Croft:2000hs}.
It has been shown, however, that the resultant constraint on neutrino
mass can vary significantly from $\sum m_\nu < 0.2$eV to $0.4$eV
depending on the specific Lyman-$\alpha$ analysis used \cite{Goobar:2006xz}.
The complication arises because the result suffers from the
systematic uncertainty regarding to the model for the intergalactic
physical effects, i.e., damping wings, ionizing radiation fluctuations,
galactic winds, and so on \cite{McDonald:2004xp}.
Therefore, we conservatively omit the Lyman-$\alpha$ forest data from
our analysis.

Because there are many other cosmological parameters than the MaVaNu
parameters, we follow the Markov Chain Monte Carlo(MCMC) global fit
approach \cite{MCMC} to explore the likelihood space and marginalize
over the nuisance parameters to obtain the constraint on
parameter(s) we are interested in. Our parameter space consists of
\begin{equation}
\vec{P}\equiv (\Omega_bh^2,\Omega_ch^2,H,\tau,A_s,n_s,m_i,\alpha,\beta)~,
\end{equation}
where $\omega_bh^2$ and $\Omega_ch^2$ are the baryon and CDM densities
in units of critical density, $H$ is the hubble parameter, $\tau$ is the
optical depth of Compton scattering to the last scattering surface, $A_s$
and $n_s$ are the amplitude and spectral index of primordial density
fluctuations, and $(m_i,\alpha,\beta)$ are the parameters of MaVaNs
defined in section III. We have put priors on MaVaNs parameters as
$\alpha>0$, and $\beta>0$ for simplicity and saving the computational time.

\begin{figure}
    \rotatebox{0}{\includegraphics[width=0.48\textwidth]{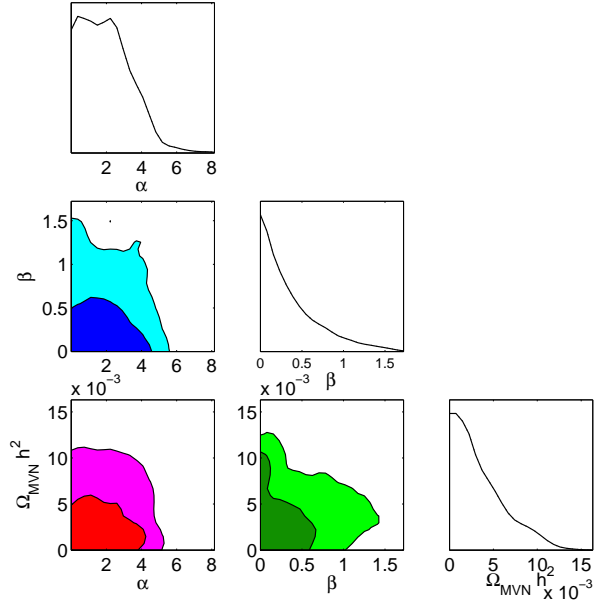}}
\caption{Contours of constant relative probabilities in two dimensional
 parameter planes for inverse power law models. Lines correspond to 68\% and 95.4\% confidence limits.}
\label{fig:ratra_tri}
\end{figure}

\begin{figure}
    \rotatebox{0}{\includegraphics[width=0.48\textwidth]{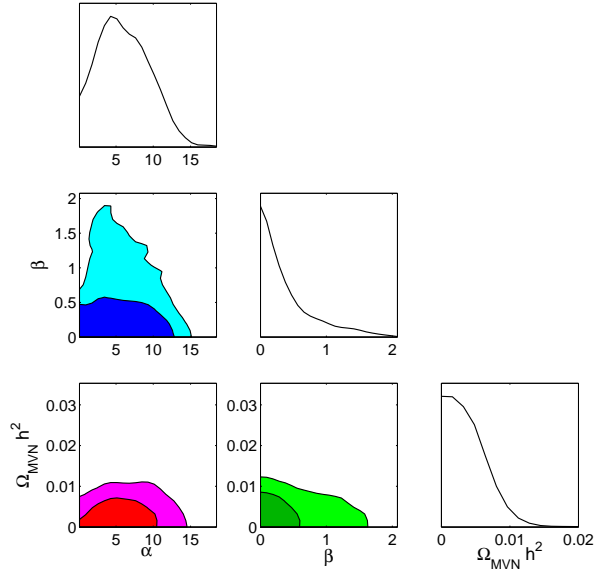}}
\caption{Same as Fig.(\ref{fig:ratra_tri}), but for SUGRA type models.
}
\label{fig:SUGRA_tri}
\end{figure}

\begin{figure}
    \rotatebox{0}{\includegraphics[width=0.48\textwidth]{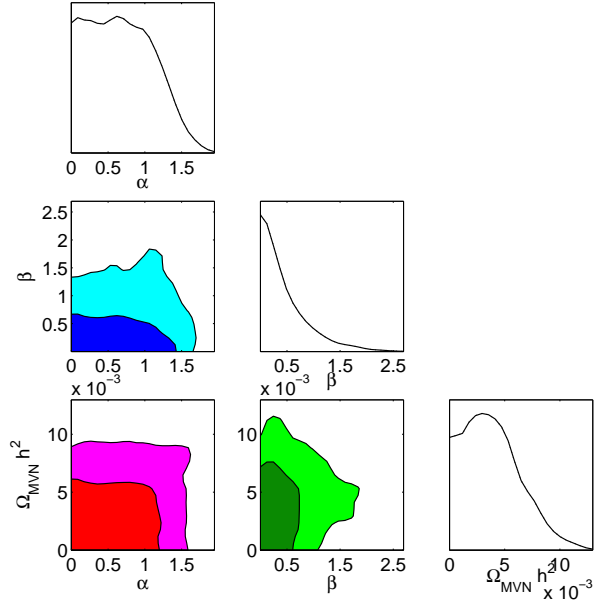}}
\caption{Same as Fig.(\ref{fig:ratra_tri}), but for exponential type models.
}
\label{fig:EXP_tri}
\end{figure}

Our results are shown in Figs.(\ref{fig:ratra_tri}) -
(\ref{fig:EXP_tri}).
In these figures we do not observe the strong degeneracy between the
introduced parameters. This is why one can put tight constraints on
MaVaNs parameters from observations. For both models we consider, larger
$\alpha$ leads larger $w$ at present. Therefore large
$\alpha$ is not allowed due to the same reason that larger $w$ is not
allowed from the current observations.

On the other hand, larger $\beta$ will generally lead larger $m_\nu$ in
the early universe. This means that the effect of neutrinos on the
density fluctuation of matter becomes larger leading to the larger
damping of the power at small scales. A complication arise because the
mass of neutrinos at the transition from the ultra-relativistic regime
to the non-relativistic one is not a monotonic function of $\beta$ as
shown in Fig.(\ref{fig:mass_evolution}). Even so, the coupled neutrinos give larger decrement
of small scale power, and therefore one can limit the coupling parameter
from the large scale structure data.

One may wonder why we can get such a tight constraint on $\beta$,
because it is naively expected that large $\beta$ value should be
allowed if $\Omega_\nu h^2 \sim 0$. In fact, a goodness of fit is still
satisfactory with large $\beta$ value when $\Omega_\nu h^2 \sim 0$.
 However, the parameters which give us the best goodness of fit
does not mean the most likely parameters in general. In our parametrization, the
accepted total volume by MCMC in the parameter space where $\Omega_\nu h^2
\sim 0$ and $\beta\ga 1$ was small, meaning that the probability of such a
parameter set is low.

\begin{center}
\begin{figure}
\vspace{0cm}
\epsfxsize=5cm
\centerline{
\rotatebox{-90}{\includegraphics[width=0.40\textwidth]{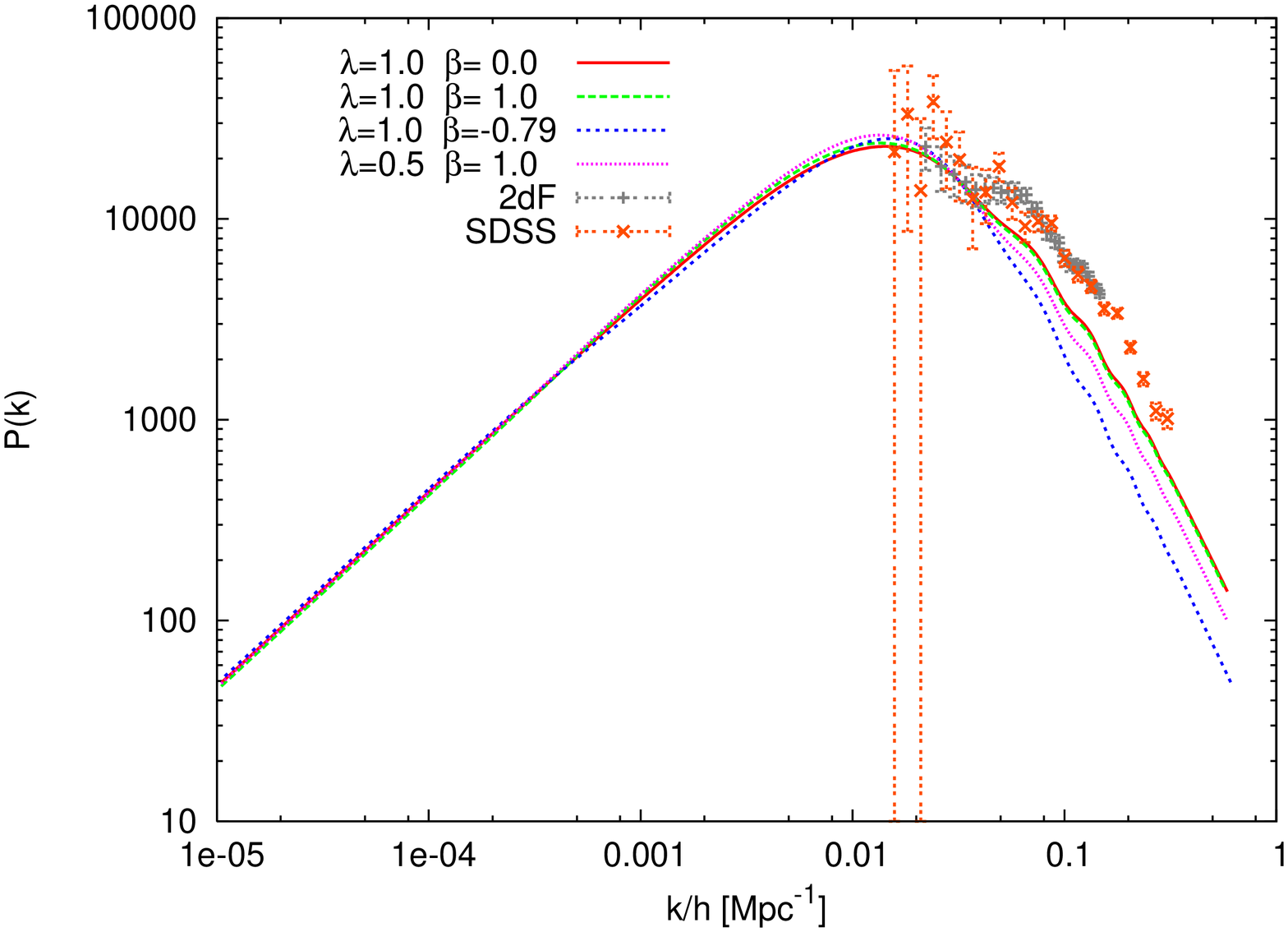}}
\rotatebox{-90}{\includegraphics[width=0.40\textwidth]{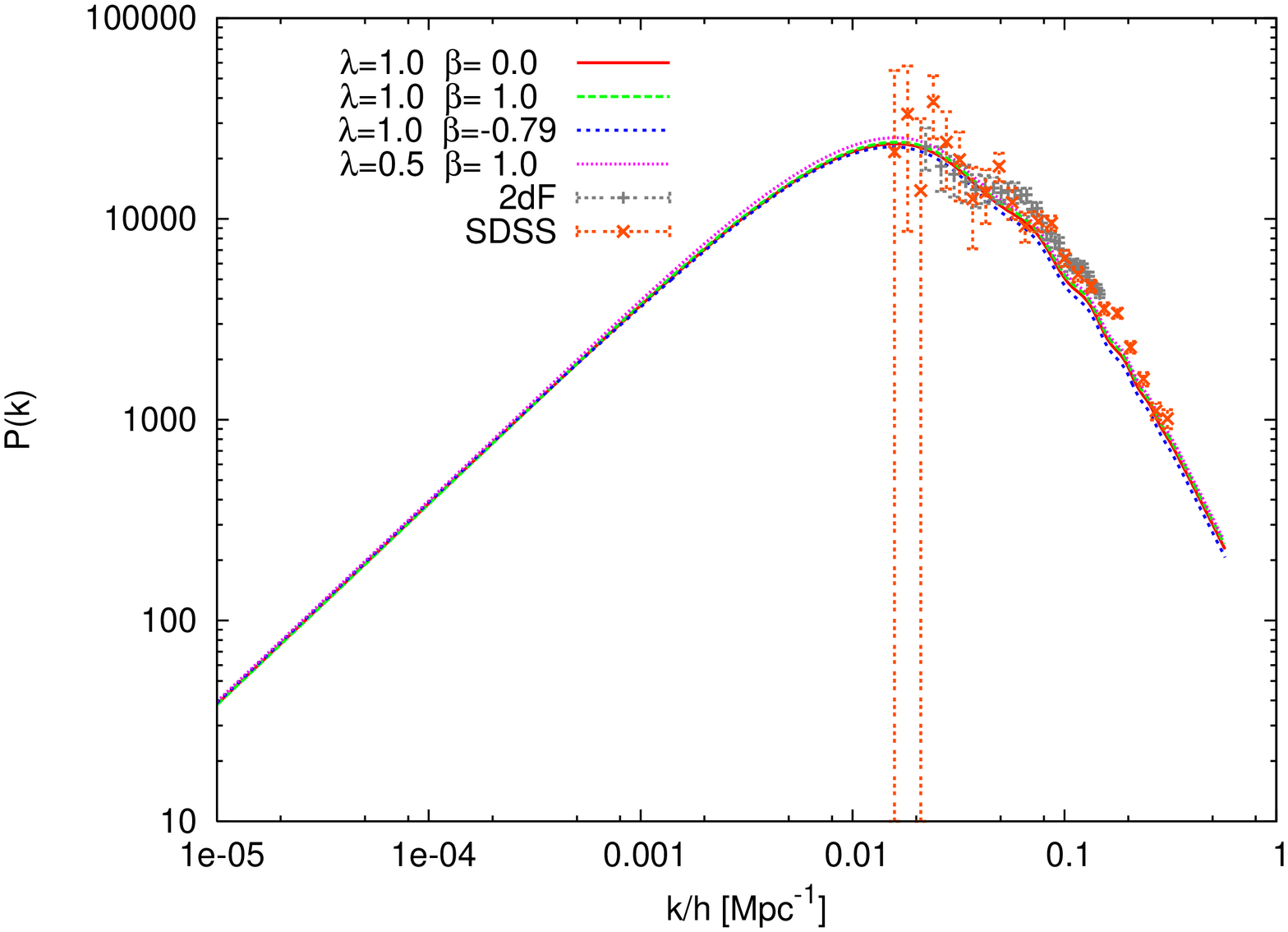}} }
\epsfxsize=5cm
\caption{Examples of the total neutrino mass contributions in power spectrum
 with $M_{\nu}=0.9\, eV$(Left panel) and with $M_{\nu}=0.3\, eV$(Right panel).
Here the variable $\lambda$ is equal to $\alpha$.
 \label{fig:nu-mass-PS} }
\end{figure}
\end{center}

We find no observational signature which favors the coupling between
MaVaNs and quintessence scalar field, and obtain the upper limit on the
coupling parameter as
\begin{equation}
\beta < 1.11,~ 1.36,~ 1.53~,
\end{equation}
and the present mass of neutrinos is also limited to
\begin{equation}
\Omega_\nu h^2_{\rm{today}} < 0.0095,~ 0.0090,~ 0.0084~,%{\mbox{for Models, I, II, III cases}}~.
\end{equation}
for models I, II and III, respectively.
When we apply the relation
between the total sum of the neutrino masses $M_{\nu}$ and their
contributions to the energy density of the universe:
$\Omega_{\nu}h^2=M_{\nu}/(93.14 eV)$, we obtain the constraint on
the total neutrino mass: $M_{\nu} < 0.87 eV (95 \% C.L.)$ in the
neutrino probe dark-energy model. The total neutrino mass
contributions in the power spectrum is shown in Fig
\ref{fig:nu-mass-PS}, where we can see the significant deviation
from observation data in the case of  large neutrino masses.
%%%%%%%%%%%%%%%%%%%%%%%%%%%%%%%%%%%%%%%%%%%%%%%%%%%%%%%%%%%%%%%%%%%%%%%%%
\begin{table}[tbh]
\caption{Global analysis data within $1\sigma$ deviation for different
types of the quintessence potential.}
{\begin{tabular}{@{}c|c|c|c|c@{}} \toprule
Quantites & Model I
&Model II & Model III & WMAP-3 data ($\Lambda$CDM)  \\ \colrule
$\Omega_B\, h^2[10^2]$ & $2.21\pm 0.07$ & $2.22\pm 0.07$ & $2.21\pm 0.07$ & $2.23\pm 0.07$   \\
$\Omega_{CDM}\, h^2[10^2]$ & $11.10\pm 0.62$ & $11.10\pm 0.65$ & $11.10 \pm 0.63$ & $12.8\pm 0.8$ \\
$H_0$ & $65.97 \pm 3.61$ & $65.37\pm 3.41$ & $65.61\pm 3.26$ & $72\pm 8$  \\
$Z_{re}$ & $10.87 \pm 2.58$ & $10.89 \pm 2.62$ & $11.07 \pm 2.44$ &   ---     \\
$\alpha$ & $< 2.63$ & $< 7.78$ & $< 0.92$ & ---       \\
$\beta$ & $< 0.46 $ & $< 0.47$ & $< 0.58$ & ---       \\
$n_s$ & $0.95\pm 0.02$ & $0.95\pm 0.02$ & $0.95\pm 0.02$ & $0.958\pm 0.016$      \\
$A_s[10^{10}]$ & $20.66\pm 1.31$ & $20.69\pm 1.32$ & $20.72\pm 1.24$ & ---- \\
$\Omega_{Q}[10^2]$ & $68.54\pm 4.81$ & $67.90\pm 4.47$ & $68.22\pm 4.17$ &
$71.6\pm 5.5$      \\
$Age/Gyrs$ & $13.95\pm 0.20$ & $13.97\pm 0.19$ & $13.69\pm 0.19$ &  $13.73\pm 0.16$ \\
$\Omega_{MVN}\,h^2[10^2]$ & $< 0.44$ & $< 0.48$ & $< 0.48$ &  $< 1.97 (95\% C.L.)$ \\
$\tau$ & $0.08\pm 0.03$ & $0.08 \pm 0.03$ & $0.09 \pm 0.03$ & $0.089 \pm 0.030$  \\ \botrule
\end{tabular} \label{ta1}}
\end{table}

%%%%%%%%%%%%%%%%%%%%%%%%%%%%%%%%%%%%%%%%%%%%%%%%%%%%%%%%%%%%%%%%%%%%%%%%%%

%%%%%%%%%%%%%%%%%%%%%%%%%%%%%%%%%%%%%%%%%%%%%%%%%%%%%%%%%%%%%
\section{Summary and Conclusion}
Before concluding the paper we should comment two important points of this paper:
 the impact of the scattering term of the Boltzmann Equation in Sec.III and 
 on the stability issue in the present models. 

\begin{center}
\begin{figure}[ht]
\vspace{0cm} \epsfxsize=5cm \centerline{
\rotatebox{0}{\includegraphics[width=0.6\textwidth]{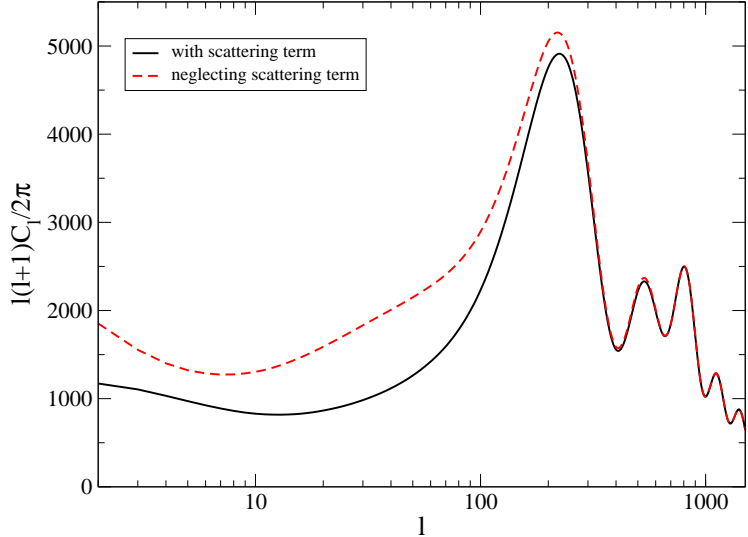}} }
\epsfxsize=5cm \caption{Differences between the CMB power spectra
with and without the scattering term in the geodesic equation of
neutrinos with the same cosmological parameters.
\label{fig:comparison-cmb-spectrum} }
\end{figure}
\end{center}

Recently, perturbation equations for the MaVaNs models were
nicely presented by Brookfield et al. \cite{Brookfield-b}, (see also
\cite{Zhao:2006zf}) which are necessary to compute CMB and LSS
spectra. A main difference here from their works is that we correctly
take into account the scattering term in the geodesic equation of
neutrinos, which was omitted there (see, however, \cite{Brookfield:2005bz}).
Because the term is proportional to
$\frac{\partial m}{\partial x}$ and first order quantity in
perturbation,  our results  and those of earlier works
\cite{Brookfield-b,Zhao:2006zf} remain the same in the background 
evolutions. However, as will be shown in the appendix, neglecting
this term violates the energy momentum conservation law at linear
level leading to the anomalously large ISW effect. Because the term
becomes important when neutrinos become massive, the late time ISW
is mainly affected through the interaction between dark energy and
neutrinos. Consequently, the differences show up at large angular
scales. In Fig. (\ref{fig:comparison-cmb-spectrum}), the differences
are shown with and without the scattering term. The early ISW can
also be affected by this term to some extent in some massive
neutrino models and the height of the first acoustic peak could be
changed. However, the position of the peaks stays almost unchanged
because the background expansion histories are the same.
%%%%%%%%%%%%%%%%%%%%%%%%%%%%%%%%%%%%%%%%%%%%
%%%%%%%%%%%%%%%%%%%%%%%%%%%%%%%%%%%%%%%%%%%%%%%%%%%%%%%%%%%%%

\begin{figure}[ht]
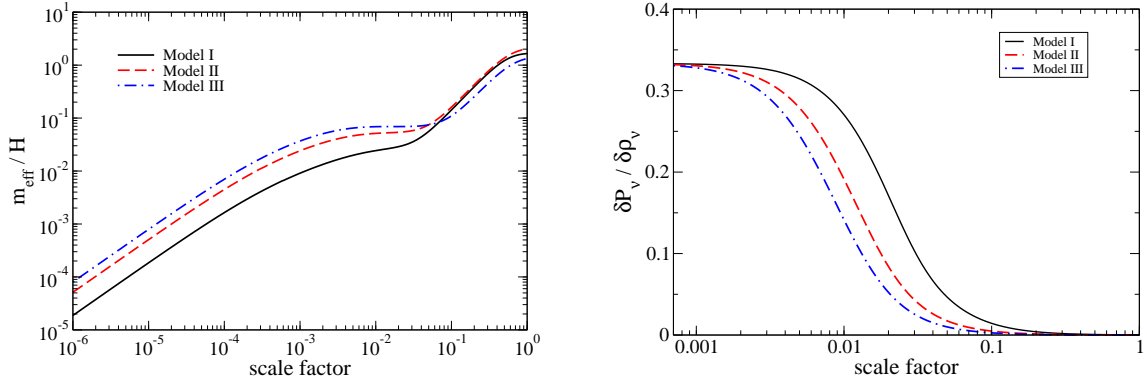

\begin{minipage}[m]{0.48\linewidth}
    \rotatebox{0}{\includegraphics[width=0.9\textwidth]{meffoH.eps}}
\end{minipage}
\begin{minipage}[m]{0.48\linewidth}
    \rotatebox{0}{\includegraphics[width=0.9\textwidth]{cs2_nu.eps}}
\end{minipage}
\caption{(Left panel): Typical evolution of the effective mass of the
 quintessence scalar field relative to theHubble scale, for all models
 considered in this paper. (Right panel): Typical evolution of the sound speed
 of neutrinos $c^s = \delta P_\nu/\delta\rho_\nu$ with the wavenumber
 $k=2.3\times 10^{-3}$ Mpc$^{-1}$, for models as
 indicated. The values stay 
 positive stating from $1/3$ (relativistic) and neutrinos are stable
 against the density fluctuation. 
}
\label{fig:m_eff}
\end{figure}

As shown in \cite{Afshordi:2005ym,Bean:2007ny}, some class of models with mass
varying neutrinos suffers from the adiabatic instability at the
first order perturbation level. This is caused by an additional force
on neutrinos mediated by the quintessence scalar field and occurs when
its effective mass is much larger than the hubble horizon scale, where
the effective mass is defined by $m_{\rm eff}^2=\frac{d^2 V_{\rm
eff}}{d\phi^2}$.  To
remedy this situation one should consider an appropriate quintessential
potential which has a mass comparable the horizon scale at present, and
the models considered in this paper are the case
\cite{Brookfield-b}. Interestingly, some authors have found that one can
construct viable MaVaNs models by choosing certain couplings and/or
quintessential potentials
\cite{Takahashi:2006jt,Kaplinghat:2006jk,Bjaelde:2007ki}. Some 
of these models even realises $m_{\rm eff} \gg H$. In  
Fig.(\ref{fig:m_eff}), masses of the scalar  
field relative to the horizon scale $m_{\rm eff}/H$ are plotted. We find that $m_{\rm eff}<H$ for almost all period and
the models are stable. We also dipict in Fig.(\ref{fig:m_eff}) the sound
speed of neutrinos defined by $c^2_s = \delta P_\nu/\delta\rho_\nu$ with
a wavenumber $k=2.3\times 10^{-3}$ Mpc$^{-1}$. 

%%%%%%%%%%%%%%%%%%%%%%%%%%%%%%%%%%%%%%%%%%%%%%%%%%%%%%%%%%%%%%%%%%%%%%%%%

In summary, we investigate dynamics of dark energy in mass-varying
neutrinos. We show and discuss many interesting aspects of the
interacting dark-energy with neutrinos scenario: (1) To explain the
present cosmological observation data, we don't need to tune the
coupling parameters between neutrinos and quintessence field, (2)
Even with a inverse power law potential or exponential type
potential which seem to be ruled out from the observation of
$\omega$ value, we can receive that the apparent value of the
equation of states can pushed down lesser than -1, (3) As a
consequence of global fit, the cosmological neutrino mass bound
beyond $\Lambda CDM$ model was first obtained with the value $\sum
m_{\nu} < 0.87 \, eV (95 \% CL)$.
More detail discussions and theoretical predictions on the equation of state
and on the absolute mass bound of neutrinos from beta decays and cosmological constrains will appear in the separated paper \cite{yyk-ichiki:2008a}.

%%%%%%%%%%%%%%%%%%%%%%%%%%%%%%%%%%%%%%%%%%%%%%%%%%%%%%%%%%%
\appendix
\section{consistency check}
The form of $\kappa$ can be also obtained by demanding
conservations of energy and momentum, i.e., demanding that $\nabla_\mu
\st{$\phi$}{\delta T^\mu_{\nu}}+\nabla_\mu \st{$\nu$}{\delta
T^\mu_{\nu}}~=~0$.
Let us begin by considering the divergence of the perturbed
stress-energy tensor for the scalar field,
\begin{eqnarray}
 \nabla_\mu \st{$\phi$}{\delta T^\mu_{\nu}} &=&
  -a^{-2}\left(\ddot{\phi}+2{\cal
      H}\dot{\phi}+a^2\frac{dV}{d\phi}\right) \partial_\nu
  \delta\phi \nonumber \\
&&-a^{-2}\left(\ddot{\delta\phi} +2{\cal
      H}\dot{\delta\phi}+k^2\delta\phi + a^2 \frac{d^2
      V}{d\phi^2}\right)\partial_\nu\phi\nonumber \\
&=&\delta\left(\frac{dV_I}{d\phi}\right)\partial_\nu{\phi}+\frac{dV_I}{d\phi}\partial_\nu{\delta\phi}
\label{eq:scalar_field_divergence}
\end{eqnarray}
where in the last line we used eqs.(\ref{eq:Qddot}) and (\ref{eq:dQddot}).
The divergence of the perturbed stress-energy tensor for the neutrinos
is given by,
\begin{equation}
\nabla_\mu \st{$\nu$}{\delta T^\mu_{0}} =
 -\dot{\delta\rho}-(\rho+P)\partial_i v_i-3{\cal H}(\delta\rho +\delta
 P)-\frac{1}{2}\dot{h}(\rho+P)
\label{eq:neutrino_time_divergence}
\end{equation}
for the time component and
\begin{equation}
\nabla_\mu \st{$\nu$}{\delta T^\mu_{i}} =
 (\rho+P)\dot{v_i}+(\dot{\rho}+\dot{P})v_i +4{\cal
 H}(\rho+P)v_i+\partial_i P +\partial_j \Sigma^j_i
\label{eq:neutrino_spatial_divergence}
\end{equation}
for the spatial component.
Let us check the energy flux conservation for example, starting with the
energy flux in neutrinos (in $k$-space):
\begin{equation}
(\rho_\nu+P_\nu)\theta_\nu=4\pi k a^{-4}\int q^2 dq qf_0(q)\Psi_1
\end{equation}
where $\theta_\nu = i k^i v_{\nu~i}$. Differentiate with respect to
$\tau$, we obtain,
\begin{eqnarray}
(\rho_\nu+P_\nu)\dot\theta_\nu+(\dot\rho_\nu+\dot P_\nu)\theta_\nu
&=& 4\pi k a^{-4}\int q^2dq q f_0 \dot \Psi_1 \nonumber \\
&&-4{\cal H} (\rho_\nu+P_\nu)\theta_\nu
\label{eq:theta_nu_dot}
\end{eqnarray}
Let us consider the first term in the right hand side of the above
equation. This gives
\begin{eqnarray}
&&4\pi k a^{-4}\int q^2dq q f_0 \dot \Psi_1 \nonumber \\
&& = 4\pi k a^{-4} \int q^2 dq q
 f_0 \left[ \frac{1}{3}\frac{q}{\epsilon}k(\Psi_0-2\Psi_2)
      +\kappa\right] \nonumber \\
&&= k^2 \delta P_\nu -k^2(\rho_\nu +P_\nu)\sigma_\nu +\frac{1}{3} 4\pi
 k^2 a^{-4}\int q^2 dq \frac{q^2}{\epsilon^2}\frac{\partial
 \epsilon}{\partial \phi}\delta\phi f_0\nonumber \\
&&~~+ 4\pi k a^{-4}\int q^2 dq q f_0 \kappa \nonumber
\end{eqnarray}
where $\sigma$ is defined as $(\rho+P)\sigma=-(k_i k_j
-\frac{1}{3}\delta_{ij})\Sigma^i_j$ and expressed by the distribution
function as
\begin{equation}
(\rho_\nu+P_\nu)\sigma_\nu =\frac{8\pi}{3}a^{-4}\int q^2 dq \frac{q^2}{\epsilon}f_0(q)\Psi_2
\end{equation}
Comparing eq. (\ref{eq:theta_nu_dot}) with eq.(\ref{eq:neutrino_spatial_divergence}), we find that
the divergence of the perturbed stress-energy tensor in spatial part for
the neutrinos leads to
\begin{equation}
\partial^i \nabla_\mu \st{$\nu$}{\delta T^\mu_{i}} = \frac{1}{3} 4\pi
 k^2 a^{-4}\int q^2 dq \frac{q^2}{\epsilon^2}\frac{\partial
 \epsilon}{\partial \phi}\delta\phi f_0 +4\pi k a^{-4} \int
 q^2 dq q f_0 \kappa
\end{equation}
On the other hand, the divergence of the perturbed stress-energy tensor
in spatial part for scalar field is, from
eq.(\ref{eq:scalar_field_divergence}),
\begin{eqnarray}
\partial^i \nabla_\mu \st{$\phi$}{\delta T^\mu_i} &=& -k^2
 \delta\phi\left(\frac{\partial \ln m_\nu}{\partial
        \phi}\right)(\rho_\nu-3P_\nu) \nonumber \\
 &=& -4\pi k^2 \delta\phi a^{-4} \int q^2 dq \frac{\partial
  \epsilon}{\partial \phi}f_0~.
\end{eqnarray}
These two equations imply that $\kappa$ shold take the form as
eq. (\ref{eq:kappa}).

Next let us check the energy conservation. Density perturbation in
neutrino is, (see eq.(\ref{eq:delta_rho_nu}))
\begin{equation}
\delta\rho_\nu=a^{-4}\int\frac{d^3 q}{(2\pi)^3} \epsilon
 f_0(q)\Psi_0+a^{-4}\int\frac{d^3 q}{(2\pi)^3}\frac{\partial
 \epsilon}{\partial \phi}\delta\phi f_0~,
\end{equation}
By differenciate with respect to $\tau$, we obtain
\begin{eqnarray}
\delta\dot\rho_\nu &=& -4{\cal H}\delta\rho_\nu + a^{-4}\int\frac{d^3
 q}{(2\pi)^3} \dot{\epsilon}f_0 \Psi_0+a^{-4}\int\frac{d^3 q}{(2\pi)^3}
 \epsilon f_0 \dot\Psi_0 \nonumber \\
&&+a^{-4}\int\frac{d^3 q}{(2\pi)^3} \frac{\partial}{\partial
 \tau}\left(\frac{\partial \epsilon}{\partial \phi}\right)\delta\phi
 f_0\nonumber \\
&&+a^{-4}\int\frac{d^3 q}{(2\pi)^3}\frac{\partial \epsilon}{\partial \phi}\dot{\delta\phi}f_0
\end{eqnarray}
where
\begin{eqnarray}
\dot\epsilon &=& ({\cal H}a^2 m^2 +a^2 m^2 \frac{\partial \ln
m_\nu}{\partial \phi}\dot\phi)/\epsilon~, \nonumber \\
\frac{\partial}{\partial\tau}\left(\frac{\partial \epsilon}{\partial
                  \phi}\right) &=& -{\cal
H}\frac{a^2m^2}{\epsilon^2}\frac{\partial \epsilon}{\partial
\phi}+2{\cal H}\frac{\partial \epsilon}{\partial \phi}+\frac{\partial^2
\epsilon}{\partial \phi^2}\dot\phi
\end{eqnarray}
Inserting eq.(\ref{eq:dot_Psi_0})
for $\dot\Psi_0$ in the above equation, we obtain
\begin{eqnarray}
\delta\dot\rho_\nu &=& -3{\cal H}(\delta\rho_\nu+\delta
 P_\nu)-(\rho_\nu+P_\nu)\theta_\nu -\frac{1}{2}\dot{h}
 (\rho_\nu+P_\nu) \nonumber \\
&&+a^{-4}\int\frac{d^3 q}{(2\pi)^3} f_0 \left(\frac{\partial^2
                     \epsilon}{\partial
                     \phi^2}\delta\phi +\Psi_0
                     \frac{\partial
                     \epsilon}{\partial
                     \phi}\right)\dot\phi \nonumber
\\
&&+a^{-4}\int\frac{d^3 q}{(2\pi)^3}f_0 \frac{\partial \epsilon}{\partial \phi}\dot{\delta\phi}
\end{eqnarray}
Comparing with eq.(\ref{eq:neutrino_time_divergence}), we find
\begin{eqnarray}
\nabla_\mu \st{$\nu$}{\delta T^\mu_{0}} &=&-a^{-4}\int\frac{d^3
 q}{(2\pi)^3} f_0 \left(\frac{\partial^2
                     \epsilon}{\partial
                     \phi^2}\delta\phi +\Psi_0
                     \frac{\partial
                     \epsilon}{\partial
                     \phi}\right)\dot\phi \nonumber
\\
&&-a^{-4}\int\frac{d^3 q}{(2\pi)^3}f_0 \frac{\partial \epsilon}{\partial \phi}\dot{\delta\phi}~,
\end{eqnarray}
which is found to be equal to $-\nabla_\mu \st{$\phi$}{\delta
T^\mu_{0}}=-\delta\left(\frac{d
V_I}{d\phi}\right)\dot\phi-\frac{dV_I}{d\phi}\delta\dot\phi$.

%\section{Boltzman Equations in Interacting Dark Energy-Neutrinos Scenario}
%\vspace{100mm}
\section{Boltzman Equations in Interacting Dark Energy-Neutrinos Scenario}
From the lagrangian ${\cal L} = m(\phi) \sqrt{-g_{\mu\nu}
\dot{x_{\mu}} \dot{x_{\nu}} }$, the Euler-Lagrange equation is given
by
\begin{equation}
{d \over d\lambda}\left(\partial {\cal L} \over \partial
\dot{x}^{\mu} \right) = {\partial {\cal L} \over \partial x^{\mu}}
\label{eq:euler-lagrange}
\end{equation}
where
\begin{eqnarray}
{\partial{\cal L} \over \partial \dot{x^{\mu}}} &=&
 P_{\mu} = -m(x^{\mu})\, { \dot{x}^{\mu} \over
 \sqrt{-g_{\alpha\beta} \dot{x}^{\alpha} \dot{x}^{\beta} } }, \\
{\partial{\cal L} \over \partial x^{\mu}} &=& {\partial m \over
\partial x^{\mu}} \sqrt{-g_{\alpha\beta} \dot{x}^{\alpha}
\dot{x}^{\beta} }   - m(x^{\mu}) {g_{\alpha\beta,\mu}
\dot{x}^{\alpha} \dot{x}^{\beta}  \over 2 \sqrt{-g_{\alpha\beta}
\dot{x}^{\alpha} \dot{x}^{\beta} } }
\end{eqnarray} Therefore eq(\ref{eq:euler-lagrange}) becomes
\begin{equation}
{1 \over \sqrt{-g_{\alpha\beta} \dot{x}^{\alpha} \dot{x}^{\beta}}}
{d \over d \lambda} \left(-m(x^{\mu})\, { \dot{x}^{\mu} \over
 \sqrt{-g_{\alpha\beta} \dot{x}^{\alpha} \dot{x}^{\beta} } } \right)
 - {m(x^{\mu}) \over 2} {g_{\alpha\beta,\mu} \dot{x}^{\alpha}
 \dot{x}^{\beta} \over g_{\alpha\beta} \dot{x}^{\alpha}
 \dot{x}^{\beta} } = {\partial m \over \partial x^{\mu}}
 \label{eq:b3}
\end{equation}
By using the relation $ds = - \sqrt{-g_{\alpha\beta} \dot{x}^{\mu}
\dot{x}^{\nu}} d\lambda$, we obtain
\begin{equation}
P^{\mu} = -m(x^{\mu}) { \dot{x}^{\mu} \over \sqrt{-g_{\alpha\beta}
\dot{x}^{\alpha} \dot{x}^{\beta} } }  = m(x^{\mu}) {dx^{\mu} \over
ds}
\end{equation}
and eq.(\ref{eq:b3}) becomes
\begin{eqnarray}
{d \over ds} \left(-m(x^{\mu}) g_{\mu\beta} {dx^{\beta}  \over ds}
\right) + {m(x^{\mu}) \over 2} g_{\alpha\beta,\mu} {dx^{\alpha}
\over ds}
{dx^{\beta} \over ds} &=& {\partial m \over \partial x^{\mu}}, \\
\cr
 {d \over ds}(g_{\mu\beta} P^{\beta}) - {1 \over 2}
 g_{\alpha\beta,\mu} P^{\alpha} {dx^{\beta} \over ds} &=& - {\partial
 m \over \partial x^{\mu}}
\end{eqnarray}
With simple calculation, finally we obtain the relations:
\begin{eqnarray}
{d P^{\nu} \over ds} + \Gamma^{\nu}_{\alpha\beta} \, P^{\alpha} { d
x^{\beta} \over ds} &=& - g^{\nu\mu} {\partial m \over \partial
x^{\mu}} \\
P^0 {d P^{\nu} \over d\tau} + \Gamma^{\nu}_{\alpha\beta} \,
P^{\alpha} P^{\beta} &=& - m g^{\nu\mu} m_{,\nu}. \label{eq:b9}
\end{eqnarray}
For $\mu=0$ component, eq.(\ref{eq:b9}) can be expressed as
\begin{equation}
{1 \over 2} {d \over d\tau} (P^{0})^2 + \Gamma^0_{\alpha\beta} =-m
g^{0\mu} m_{,\nu}. \label{eq:b10}
\end{equation}
Since $P^0 = g^{00} P_0= a_2 \epsilon$, each terms of the
eq.(\ref{eq:b10}) are given by:
\begin{eqnarray}
{\rm First \,\, term} &=& -2 a^{-4} H q^2 + a^{-4} q
{dq \over d\tau} - a^{-2} H m^2 + a^{-2} m {dm \over d\tau} \\
{\rm Second \,\, term} &=& 2 a^{-4} H q^2 + a^{-2} H m^{2} + a^{-4}
{1 \over 2} \dot{h}_{ij} q^{i} q^{j} \\
{\rm Third \,\, term} &=& a^{-2} m {\partial m \over \partial\tau}
\end{eqnarray}
Since the first term includes the total derivative w.r.t. comoving
time, we obtain finally the eq.(\ref{eq:eq-b}) in Section III-C:
\begin{equation}
{dq \over d\tau} = -{1 \over 2} \dot{h}_{ij}\, q \, n^{i} n^{j} -
a^2 {m \over q} {\partial m \over \partial x^{i}} \, {d x^{i} \over
d \tau}.
\end{equation}
%%%%%%%%%%%%%%%%%%%%%%%%%%%%%%%%%%%%%%%%%%%%%%%

%\section{Mass-Varying Neutrinos in High-redshift region}
%\vspace{100mm}

%\appendix

%  \begin{figure}[h]
%    \rotatebox{0}{\includegraphics[width=0.48\textwidth]{exponential.eps}}
%    \rotatebox{0}{\includegraphics[width=0.48\textwidth]{exponential_mass.eps}}
%    \rotatebox{0}{\includegraphics[width=0.48\textwidth]{exponential_cl.eps}}
%  \caption{
%  $V=M^4 \exp(-3.7(Q/M_{\rm pl}))$, $M=M_i \exp(20(Q/M_{\rm pl})).$
%  }
%  \end{figure}
%  \begin{figure}[h]
%    \rotatebox{0}{\includegraphics[width=0.48\textwidth]{ratra_a5_b5.eps}}
%    \rotatebox{0}{\includegraphics[width=0.48\textwidth]{Ratra_a5_b5_mass.eps}}
%    \rotatebox{0}{\includegraphics[width=0.48\textwidth]{Ratra_a5_b5_cls.eps}}
%  \caption{
%  $V=M^4 \frac{M^5}{\phi^5}$, $M=M_i \exp(\phi/M_{\rm pl}).$
%  }
%  \end{figure}

%  \begin{figure}[h]
%    \rotatebox{0}{\includegraphics[width=0.48\textwidth]{ratra_matters.eps}}
%    \rotatebox{0}{\includegraphics[width=0.48\textwidth]{sugra_matters.eps}}
%    \rotatebox{0}{\includegraphics[width=0.48\textwidth]{SUGRA_a2_b1_Cls.eps}}
%  \caption{
%  matter power spectra.
%  }
%  \end{figure}
\acknowledgements We would like to thank L. Amendola, O. Seto, S. Carroll,
and L. Schrempp  
for useful comments and discussion. 
K.I. thanks C.~van de Bruck for useful communications.
K.I. thanks also to KICP and National Taiwan university for kind hospitalities
where most part of this work have been done. K.I.'s work is
supported by Grant-in-Aid for JSPS Fellows. Y.Y.K's work is
partially supported by Grants-in-Aid for NSC in Taiwan, Center
for High Energy Physics(CHEP)/KNU and APCTP in Korea.

\end{document}